\documentclass[twocolumn,preprintnumbers]{revtex4}%
\usepackage{amssymb}
\usepackage{amsmath}
\usepackage{graphicx}
\usepackage{dcolumn}
\usepackage{bm}
\usepackage{xcolor}
\usepackage{ifthen}
\usepackage[normalem]{ulem}
\usepackage{amsfonts}%
\UseRawInputEncoding
\providecommand{\U}[1]{\protect\rule{.1in}{.1in}}
\providecommand{\U}[1]{\protect\rule{.1in}{.1in}}
\def\showal{1}
\newcommand{\al}[1]{\ifthenelse{\showal=1}{\textcolor{orange}{[[#1]]}}{}}

\newcommand{\eb}[1]{\ifthenelse{\showal=1}{\textcolor{cyan}{[[#1]]}}{}}
\begin{document}
\title{Thermalization and disentanglement with a nonlinear Schr\"{o}dinger equation}
\author{Eyal Buks}
\email[]{eyal@ee.technion.ac.il}
\affiliation{Andrew and Erna Viterbi Department of Electrical Engineering, Technion, Haifa
32000, Israel}
\date{\today }

\begin{abstract}
We study a recently proposed modified Schr\"{o}dinger equation having an added
nonlinear term. For the case where a stochastic term is added to the
Hamiltonian, the fluctuating response is found to resemble the process of
thermalization. Disentanglement induced by the added nonlinear term is
explored for a system made of two coupled spins. A butterfly-like effect is
found near fully entangled states of the spin-spin system. A limit cycle
solution is found when one of the spins is externally driven.

\end{abstract}
\maketitle





\section{Introduction}

\label{SecIntro}

In 1935 \cite{Schrodinger_807} Schr\"{o}dinger has identified a
self-inconsistency in the quantum to classical transition process
\cite{Penrose_4864,Leggett_939,Leggett_022001}, which became known as the
problem of quantum measurement. This problem is related to the phenomenon of
quantum entanglement. Exploring possible mechanisms of disentanglement, may
help resolving this long-standing problem.

Processes such as disentanglement require nonlinear time evolution. A variety
of nonlinear terms \cite{Geller_2111_05977} that can be added to the
Schr\"{o}dinger equation have been explored before
\cite{Weinberg_336,Weinberg_61,Doebner_397,Doebner_3764,Gisin_5677,Kaplan_055002,Munoz_110503}%
. In most previous proposals, the purpose of the added nonlinear terms is to
generate a spontaneous collapse \cite{Bassi_471}.

Here we explore a recently proposed modified Schr\"{o}dinger equation having
an added nonlinear term \cite{Buks_355303} (see section \ref{SecMSE}). The
proposed equation can be constructed for any physical system having Hilbert
space of finite dimensionality, and it does not violate unitarity of the time evolution.

The effect of the added term on the dynamics of a single spin 1/2 is studied
in section \ref{SecOneS}. The spin's response to an applied fluctuating
magnetic field is found to mimic the process of thermalization
\cite{Diosi_451,Molmer_524,Dalibard_580,Semin_063313} (see section
\ref{SecSpinTE}). Disentanglement induced by the nonlinear term is explored
with two coupled spins (see section \ref{SecTwoSpins}). A butterfly-like
effect is found near fully entangled spin-spin states.

The system can become unstable when one spin is externally driven (see section
\ref{SecInstab}). Limit cycle solutions for the modified Schr\"{o}dinger
equation are found in the instability region. The instability of the modified
Schr\"{o}dinger equation is closely related to an instability found with a
similar spin-spin system \cite{Levi_053516}, when the equations of motion
generated by the standard Schr\"{o}dinger equation are analyzed using the mean
field approximation
\cite{breuer2002theory,Drossel_217,Hicke_024401,Klobus_034201}.

\section{The modified Schr\"{o}dinger equation}

\label{SecMSE}

Let $\mathcal{H}$ be a time-independent Hermitian Hamiltonian of a given
physical system. Consider a modified Schr\"{o}dinger equation for the state
vector $\left\vert \psi\right\rangle $ given by \cite{Buks_355303}%
\begin{equation}
\frac{\mathrm{d}}{\mathrm{d}t}\left\vert \psi\right\rangle =\left(
-i\hbar^{-1}\mathcal{H}+\gamma_{\mathrm{D}}M_{\mathrm{D}}\right)  \left\vert
\psi\right\rangle \;, \label{MSE}%
\end{equation}
where $\mathrm{d}/\mathrm{d}t$ is a time derivative. In the added (to the
standard\ Schr\"{o}dinger equation) term $\gamma_{\mathrm{D}}M_{\mathrm{D}}$,
the rate $\gamma_{\mathrm{D}}$ is a positive coefficient, and the operator
$M_{\mathrm{D}}$ is derived from a given non-zero state vector $\left\vert
\Psi\right\rangle $ according to (the state vector $\left\vert \Psi
\right\rangle $ is not required to be normalized)%
\begin{equation}
M_{\mathrm{D}}=-\sqrt{\frac{\left\langle \Psi\right.  \left\vert
\Psi\right\rangle }{1-\left\langle \mathcal{P}\right\rangle }}\left(
\mathcal{P}-\left\langle \mathcal{P}\right\rangle \right)  \;, \label{M_D}%
\end{equation}
where the projection operator $\mathcal{P}$ is given by%
\begin{equation}
\mathcal{P}=\frac{\left\vert \Psi\right\rangle \left\langle \Psi\right\vert
}{\left\langle \Psi\right.  \left\vert \Psi\right\rangle }\;,
\end{equation}
and the expectation value $\left\langle \mathcal{P}\right\rangle $ is given by%
\begin{equation}
\left\langle \mathcal{P}\right\rangle =\frac{\left\langle \psi\right\vert
\mathcal{P}\left\vert \psi\right\rangle }{\left\langle \psi\right.  \left\vert
\psi\right\rangle }=\frac{\left\vert \left\langle \Psi\right.  \left\vert
\psi\right\rangle \right\vert ^{2}}{\left\langle \Psi\right.  \left\vert
\Psi\right\rangle \left\langle \psi\right.  \left\vert \psi\right\rangle }\;.
\end{equation}

The modified Schr\"{o}dinger equation yields a modified master equation for
the pure state density operator $\rho=\left\vert \psi\right\rangle
\left\langle \psi\right\vert $ given by (note that $M_{\mathrm{D}}^{\dag
}=M_{\mathrm{D}}^{{}}$ and $\mathcal{H}^{\dag}=\mathcal{H}^{{}}$)%
\begin{equation}
\frac{\mathrm{d}\rho}{\mathrm{d}t}=\frac{\left[  \mathcal{H},\rho\right]
}{i\hbar}+\gamma_{\mathrm{D}}\left(  \rho M_{\mathrm{D}}+M_{\mathrm{D}}%
\rho\right)  \;. \label{MME}%
\end{equation}
Note that $\left(  \mathrm{d/d}t\right)  \operatorname{Tr}\rho=0$ provided
that $\operatorname{Tr}\rho=1$ (i.e. $\left\vert \psi\right\rangle $ is
normalized) [see Eq. (\ref{M_D}), and note that $\left\langle O\right\rangle
\equiv\operatorname{Tr}\left(  \rho O\right)  $ for an arbitrary observable
$O^{{}}=O^{\dag}$], and that $\left(  \mathrm{d/d}t\right)  \operatorname{Tr}%
\rho^{2}=0$ provided that $\rho^{2}=\rho$ [note that $\left\langle
M_{\mathrm{D}}\right\rangle =0$, see Eq. (\ref{M_D})]. Henceforth it is
assumed that $\left\vert \psi\right\rangle $ is normalized, and that $\rho
^{2}=\rho$. The modified master equation (\ref{MME}) yields a modified
Heisenberg equation given by%
\begin{equation}
\frac{\mathrm{d}\left\langle O\right\rangle }{\mathrm{d}t}=\frac{\left\langle
\left[  O,\mathcal{H}\right]  \right\rangle }{i\hbar}+\gamma_{\mathrm{D}%
}\left\langle M_{\mathrm{D}}O+OM_{\mathrm{D}}\right\rangle \;, \label{MHE}%
\end{equation}
where $O^{{}}=O^{\dag}$ is a given observable that does not explicitly depend
on time.

\section{One spin}

\label{SecOneS}

As an example, consider a spin 1/2 particle. The $2\times2$ density matrix
$\rho$ is expressed as%
\begin{equation}
\rho=\frac{1+\boldsymbol{k}\cdot\boldsymbol{\sigma}}{2}\;,
\end{equation}
where $\boldsymbol{k}=\left(  k_{x},k_{y},k_{z}\right)  $ is a real vector,
and $\boldsymbol{\sigma}=\left(  \sigma_{x},\sigma_{y},\sigma_{z}\right)  $ is
the Pauli matrix vector%
\begin{equation}
\sigma_{x}=\left(
\begin{array}
[c]{cc}%
0 & 1\\
1 & 0
\end{array}
\right)  ,\;\sigma_{y}=\left(
\begin{array}
[c]{cc}%
0 & -i\\
i & 0
\end{array}
\right)  ,\;\sigma_{z}=\left(
\begin{array}
[c]{cc}%
1 & 0\\
0 & -1
\end{array}
\right)  \;. \label{Pauli}%
\end{equation}
The Hamiltonian $\mathcal{H}$ is assumed to be given by $\hbar^{-1}%
\mathcal{H}=\boldsymbol{\omega\cdot\sigma}$, where
$\boldsymbol{\omega}=\left(  \omega_{x},\omega_{y},\omega_{z}\right)  $ is a
constant real vector. With the help of the identity $\left(
\boldsymbol{\sigma}\cdot\boldsymbol{a}\right)  \left(
\boldsymbol{\sigma}\cdot\boldsymbol{b}\right)  =\boldsymbol{a}\cdot
\boldsymbol{b}+i\boldsymbol{\sigma}\cdot\left(  \boldsymbol{a}\times
\boldsymbol{b}\right)  $, where $\boldsymbol{a}$ and $\boldsymbol{b}$ are
three-dimensional vectors, one finds that [see Eq. (\ref{MHE}), and note that
$\operatorname{Tr}\sigma_{x}=\operatorname{Tr}\sigma_{y}=\operatorname{Tr}%
\sigma_{z}=0$,]%
\begin{equation}
\frac{\mathrm{d}\boldsymbol{k}}{\mathrm{d}t}=2\left(
\boldsymbol{\omega}\times\boldsymbol{k}\right)  +\gamma_{\mathrm{D}%
}\left\langle M_{\mathrm{D}}%
\boldsymbol{\sigma}+\boldsymbol{\sigma}M_{\mathrm{D}}\right\rangle \;.
\end{equation}
Consider the case where $\left\vert \Psi\right\rangle $ is taken to be a
normalized eigenvector of $\boldsymbol{\hat{s}\cdot\sigma}$, where
$\boldsymbol{\hat
{s}}=\left(  s_{x},s_{y},s_{z}\right)  $ is a constant real unit vector, and
the corresponding eigenvalue is $+1$ (i.e. $\left\langle \Psi\right.
\left\vert \Psi\right\rangle =1$, $\boldsymbol{\hat{s}\cdot\hat{s}}=1$ and
$\boldsymbol{\hat{s}\cdot\sigma}\left\vert \Psi\right\rangle =\left\vert
\Psi\right\rangle $). For this case $\sqrt{\left(  1-\boldsymbol{\hat{s}}\cdot
\boldsymbol{k}\right)  /2}\left\langle M_{\mathrm{D}}\boldsymbol{\sigma
}+\boldsymbol{\sigma}M_{\mathrm{D}}\right\rangle =\left(
\boldsymbol{\hat{s}}\cdot\boldsymbol{k}\right)
\boldsymbol{k}-\boldsymbol{\hat{s}}$ [see Eq. (\ref{M_D})], and thus (compare
with Refs. \cite{Kowalski_1,Fernengel_385701,Kowalski_167955})%
\begin{equation}
\frac{\mathrm{d}\boldsymbol{k}}{\mathrm{d}t}=2\boldsymbol{\omega}\times
\boldsymbol{k}+\gamma_{\mathrm{D}}\frac{\left(  \boldsymbol{\hat{s}}\cdot
\boldsymbol{k}\right)  \boldsymbol{k-\hat{s}}}{\sqrt{\frac
{1-\boldsymbol{\hat{s}}\cdot\boldsymbol{k}}{2}}}\;. \label{k eom}%
\end{equation}

The following holds $\boldsymbol{\hat{s}}=\boldsymbol{\hat{s}}_{\parallel
}+\boldsymbol{\hat{s}}_{\perp}$, where the parallel
$\boldsymbol{\hat{s}}_{\parallel}$ and perpendicular
$\boldsymbol{\hat{s}}_{\perp}$ (with respect to $\boldsymbol{k}$) components
are given by $\boldsymbol{\hat{s}}_{\parallel}=\left(  \boldsymbol{k}\cdot
\boldsymbol{k}\right)  ^{-1}\left(  \boldsymbol{k}\cdot
\boldsymbol{\hat{s}}\right)  \boldsymbol{k}$ and $\boldsymbol{\hat{s}}_{\perp
}=-\left(  \boldsymbol{k}\cdot\boldsymbol{k}\right)  ^{-1}\boldsymbol{k}\times
\left(  \boldsymbol{k}\times\boldsymbol{\hat{s}}\right)  $ [recall the vector
identity $\boldsymbol{A}\times\left(  \boldsymbol{B}\times
\boldsymbol{C}\right)  =\left(  \boldsymbol{A}\cdot\boldsymbol{C}\right)
\boldsymbol{B}-\left(  \boldsymbol{A}\cdot\boldsymbol{B}\right)
\boldsymbol{C}$]. Thus, for the case where $\left\vert
\boldsymbol{k}\right\vert =1$ (i.e. $\boldsymbol{k}\cdot\boldsymbol{k}=1$) Eq.
(\ref{k eom}) becomes%
\begin{align}
\frac{\mathrm{d}\boldsymbol{k}}{\mathrm{d}t}  &  =2\boldsymbol{\omega}\times
\boldsymbol{k}-\gamma_{\mathrm{D}}\frac{\boldsymbol{\hat{s}}_{\perp}}%
{\sqrt{\frac{1-\boldsymbol{\hat{s}}_{\parallel}\cdot\boldsymbol{k}}{2}}%
}\nonumber\\
&  =\left(  2\boldsymbol{\omega}+\gamma_{\mathrm{D}}\frac
{\boldsymbol{\hat{s}}\times\boldsymbol{k}}{\sqrt{\frac
{1-\boldsymbol{\hat{s}}\cdot\boldsymbol{k}}{2}}}\right)  \times
\boldsymbol{k}\;.\nonumber\\
&  \label{k eom par pen}%
\end{align}
The above result (\ref{k eom par pen}) indicates that the radial component of
$\mathrm{d}\boldsymbol{k/}\mathrm{d}t$ vanishes [i.e. $\left(  \mathrm{d}%
\boldsymbol{k/}\mathrm{d}t\right)  \cdot\boldsymbol{k}=0$] on the surface of
the Bloch sphere (i.e. when $\left\vert \boldsymbol{k}\right\vert =1$).

\begin{figure}[b]
\begin{center}
\includegraphics[width=1.0\columnwidth]{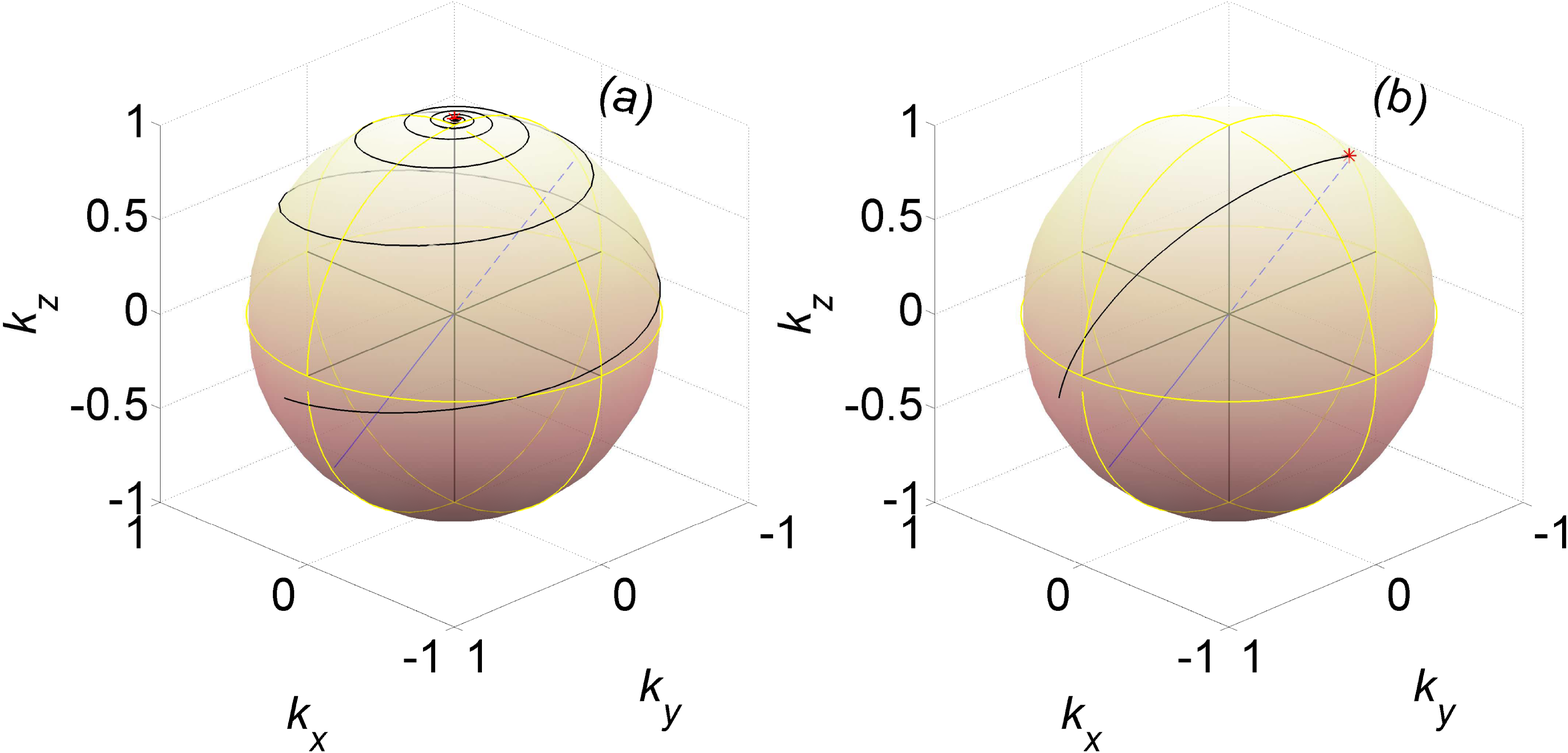}
\end{center}
\caption{One spin. The black solid line is obtained by numerically integrating
of the one-spin modified Schr\"{o}dinger equation (\ref{k eom}). The blue
solid (dashed) line connects the origin and the point $\boldsymbol{\hat{s}}$
($-\boldsymbol{\hat{s}}$), and $\boldsymbol{\omega}$ is parallel to the
$\boldsymbol{\hat
{z}}$ direction. (a) The ratio $\gamma_{\mathrm{D}}/\left\vert
\boldsymbol{\omega
}\right\vert =0.25$. The red star symbol represents the analytical
approximation for the $\boldsymbol{k}$ fixed point given by $\pm\left(
\boldsymbol{\hat{\omega}}+\left(  \gamma_{\mathrm{D}}/\left\vert
\boldsymbol{\omega
}\right\vert \right)  \left(  2\left(  1-\boldsymbol{\hat{s}}\cdot
\boldsymbol{\hat
{\omega}}\right)  \right)  ^{-1/2}\boldsymbol{\hat{s}}\times
\boldsymbol{\hat{\omega}}\right)  $, which is valid when $\gamma_{\mathrm{D}%
}\ll\left\vert \boldsymbol{\omega}\right\vert $. (b) The ratio $\gamma
_{\mathrm{D}}/\left\vert \boldsymbol{\omega}\right\vert =25$. The red star
symbol represents the analytical approximation for the $\boldsymbol{k}$ fixed
point given by $-\boldsymbol{\hat{s}}+2\left(  \left\vert
\boldsymbol{\omega}\right\vert /\gamma_{\mathrm{D}}\right)
\boldsymbol{\hat{s}}\times\boldsymbol{\hat{\omega}}$, which is valid when
$\gamma_{\mathrm{D}}\gg\left\vert \boldsymbol{\omega}\right\vert $.}%
\label{FigOneSpin}%
\end{figure}

By multiplying Eq. (\ref{k eom}) by
$\boldsymbol{\hat{\omega}}=\boldsymbol{\omega
}/\left\vert \boldsymbol{\omega}\right\vert $ one finds that%
\begin{equation}
\frac{\mathrm{d}k_{\parallel}}{\mathrm{d}t}=\gamma_{\mathrm{D}}\frac
{\boldsymbol{\hat{s}}\cdot\left(  \left(  \boldsymbol{k}\cdot
\boldsymbol{\hat{\omega}}\right)  \boldsymbol{k-\hat{\omega}}\right)  }%
{\sqrt{\frac{1-\boldsymbol{\hat{s}}\cdot\boldsymbol{k}}{2}}}\;,
\label{d k par / dt}%
\end{equation}
where $k_{\parallel}=\boldsymbol{k}\cdot\boldsymbol{\hat{\omega}}$, hence
$\mathrm{d}k_{\parallel}/\mathrm{d}t=0$ for $\boldsymbol{k}=\pm
\boldsymbol{\hat
{\omega}}$. For the case where
$\boldsymbol{\hat{\omega}}=\boldsymbol{\hat{z}}$ Eq. (\ref{d k par / dt})
becomes%
\begin{equation}
\frac{\mathrm{d}k_{\parallel}}{\mathrm{d}t}=\gamma_{\mathrm{D}}\frac{\left(
s_{x}k_{x}+s_{y}k_{y}\right)  k_{z}+s_{z}\left(  k_{z}^{2}-1\right)  }%
{\sqrt{\frac{1-\boldsymbol{\hat{s}}\cdot\boldsymbol{k}}{2}}}\;.
\label{d k par / dt om=z}%
\end{equation}
When $\gamma_{\mathrm{D}}\ll\left\vert \boldsymbol{\omega}\right\vert $ the
dynamics is dominated by the term $2\boldsymbol{\omega}\times\boldsymbol{k}$
in Eq. (\ref{k eom}), which gives rise to spin precession. For this case the
averaged value of the term $s_{x}k_{x}+s_{y}k_{y}$ in Eq. (\ref{d k par / dt om=z}) is nearly zero. Note that
$k_{z}^{2}-1\leq0$, hence for this case $\boldsymbol{k}\rightarrow
+\boldsymbol{\hat
{\omega}}$ ($\boldsymbol{k}\rightarrow-\boldsymbol{\hat{\omega}}$) in the
limit $t\rightarrow\infty$ when $s_{z}<0$ ($s_{z}>0$). The plot shown in Fig.
\ref{FigOneSpin}(a) is obtained by numerically integrating the modified
Schr\"{o}dinger equation (\ref{k eom}) for the case where
$\boldsymbol{\omega}$ is parallel to the $\boldsymbol{\hat{z}}$ direction, and
$\gamma_{\mathrm{D}}/\left\vert \boldsymbol{\omega}\right\vert =0.25$. As can
be seen from Eq. (\ref{k eom par pen}), to first order in $\gamma_{\mathrm{D}%
}/\left\vert \boldsymbol{\omega}\right\vert $ the $\boldsymbol{k}$ fixed point
is located at $\pm\left(  \boldsymbol{\hat{\omega}}+\left(  \gamma
_{\mathrm{D}}/\left\vert \boldsymbol{\omega}\right\vert \right)  \left(
2\left(  1-\boldsymbol{\hat{s}}\cdot\boldsymbol{\hat{\omega}}\right)  \right)
^{-1/2}\boldsymbol{\hat{s}}\times\boldsymbol{\hat{\omega}}\right)  $ [see the
red star symbol in Fig. \ref{FigOneSpin}(a)].

In the opposite extreme case of $\gamma_{\mathrm{D}}\gg\left\vert
\boldsymbol{\omega}\right\vert $, the dynamics is dominated by the term
proportional to $\gamma_{\mathrm{D}}$ in Eq. (\ref{k eom}). Note that $\left(
\boldsymbol{\hat{s}}\cdot\boldsymbol{k}\right)  \boldsymbol{k-\hat{s}}=0$ for
$\boldsymbol{k}=\pm\boldsymbol{\hat{s}}$, and that the term proportional to
$\gamma_{\mathrm{D}}$ in Eq. (\ref{k eom}) gives rise to attraction
(repulsion) of $\boldsymbol{k}$ to the point $-\boldsymbol{\hat{s}}$
($+\boldsymbol{\hat
{s}}$). Consider the case where
$\boldsymbol{k}=-\boldsymbol{\hat{s}}+\boldsymbol{\delta}$, and where
$\boldsymbol{\hat{s}}\cdot\boldsymbol{\delta}=0$ ($\boldsymbol{\delta}$ is
considered as infinitesimally small). To first order in $\left\vert
\boldsymbol{\delta}\right\vert $ Eq. (\ref{k eom}) yields%
\begin{equation}
\frac{\mathrm{d}\boldsymbol{\delta}}{\mathrm{d}t}=2\boldsymbol{\omega}\times
\left(  -\boldsymbol{\hat{s}}+\boldsymbol{\delta}\right)  -\gamma_{\mathrm{D}%
}\boldsymbol{\delta }+O\left(  \left\vert \boldsymbol{\delta}\right\vert
^{2}\right)  \;.
\end{equation}
hence the $\boldsymbol{k}$ point $-\boldsymbol{\hat{s}}+2\left(  \left\vert
\boldsymbol{\omega}\right\vert /\gamma_{\mathrm{D}}\right)
\boldsymbol{\hat{s}}\times\boldsymbol{\hat{\omega}}$ is nearly a stable fixed
point when $\gamma_{\mathrm{D}}\gg\left\vert \boldsymbol{\omega}\right\vert $
[see the red star symbol in Fig. \ref{FigOneSpin}(b)].

\section{Spin in thermal equilibrium}

\label{SecSpinTE}

The effect of coupling between the spin and its environment can be accounted
for using the modified Schr\"{o}dinger equation (\ref{MSE}) provided that a
fluctuating magnetic field is added \cite{Slichter_Principles}. Consider the
case where the spin Hamiltonian $\mathcal{H}$ is given by $\hbar
^{-1}\mathcal{H}=\boldsymbol{\omega\cdot\sigma}$, where
$\boldsymbol{\omega}=\omega_{0}\boldsymbol{\hat{z}}+\left(  \omega_{x}%
,\omega_{y},\omega_{z}\right)  $, where $\omega_{0}$ is a constant, and where
$\omega_{x}\left(  t\right)  $, $\omega_{y}\left(  t\right)  $ and $\omega
_{z}\left(  t\right)  $ represent the effect of a fluctuating magnetic field.
The following is assumed to hold $\left\langle \omega_{x}\right\rangle
=\left\langle \omega_{y}\right\rangle =\left\langle \omega_{z}\right\rangle
=0$, where $\left\langle {}\right\rangle $ denotes time averaging (i.e. the
fluctuating field has a vanishing averaged value), and the correlation
function $\left\langle \omega_{i}\left(  t\right)  \omega_{j}\left(
t^{\prime}\right)  \right\rangle $ is given by%
\begin{equation}
\left\langle \omega_{i}\left(  t\right)  \omega_{j}\left(  t^{\prime}\right)
\right\rangle =\delta_{ij}\omega_{\mathrm{s}}^{2}\exp\left(  -\frac{\left\vert
t-t^{\prime}\right\vert }{\tau_{\mathrm{s}}}\right)  \;, \label{omega ACF}%
\end{equation}
where both the variance $\omega_{\mathrm{s}}^{2}$ and the correlation time
$\tau_{\mathrm{s}}$ are positive constants, and where $i,j\in\left\{
x,y,z\right\}  $. The added fluctuating magnetic field gives rise to
longitudinal $T_{\mathrm{s}1}^{-1}$ and transverse $T_{\mathrm{s}2}^{-1}$
relaxation rates given by [see Eqs. (17.274) and (17.275) of Ref.
\cite{Buks_QMLN}]%
\begin{equation}
\frac{1}{T_{\mathrm{s}1}}=\frac{2\omega_{\mathrm{s}}^{2}\tau_{\mathrm{s}}%
}{1+\omega_{0}^{2}\tau_{\mathrm{s}}^{2}}\;, \label{1/T1}%
\end{equation}
and%
\begin{equation}
\frac{1}{T_{\mathrm{s}2}}=\frac{1}{2T_{\mathrm{s}1}}+\omega_{\mathrm{s}}%
^{2}\tau_{\mathrm{s}}\;. \label{1/T2}%
\end{equation}

\begin{figure}[b]
\begin{center}
\includegraphics[width=1.0\columnwidth]{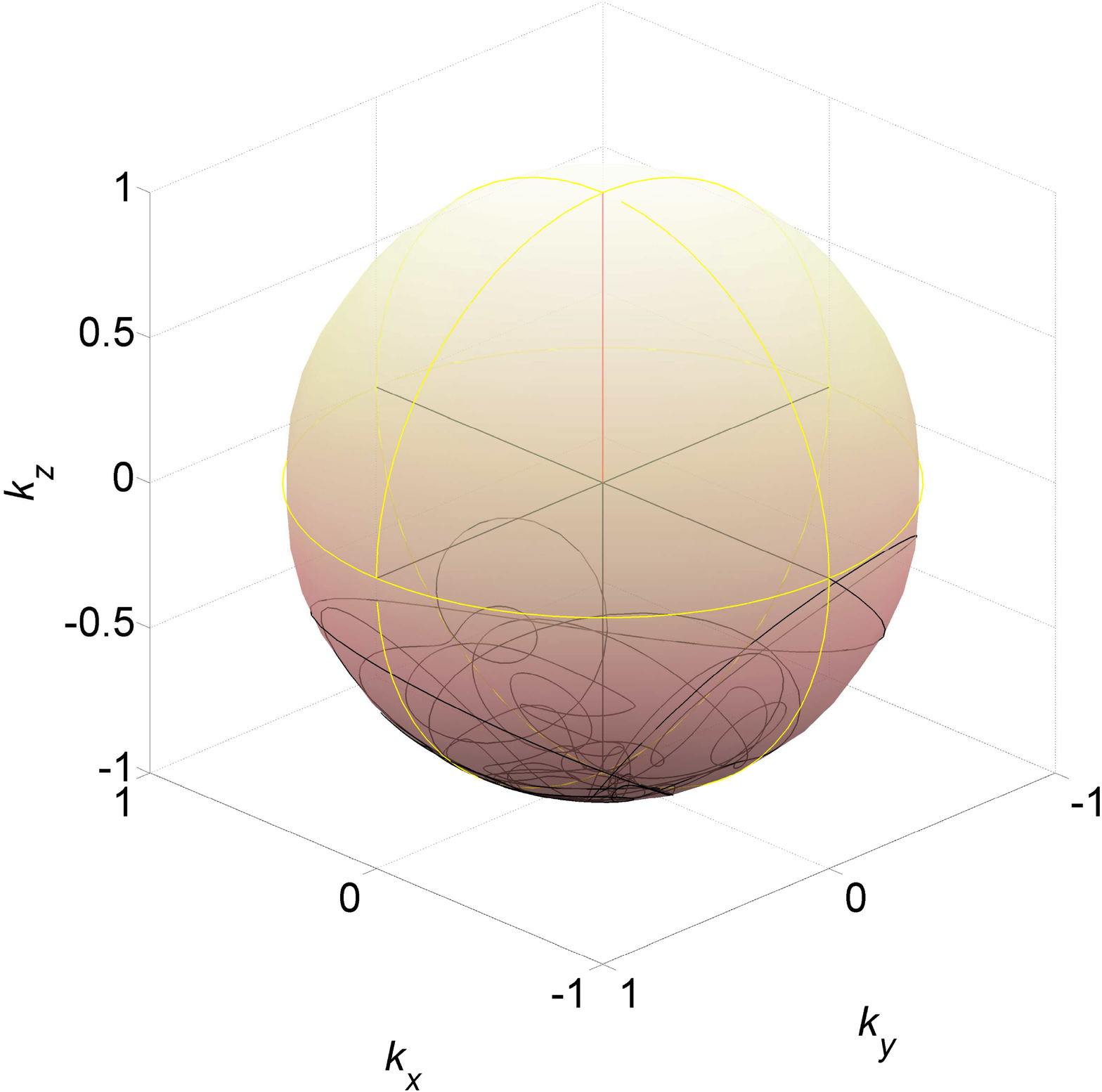}
\end{center}
\caption{Thermal equilibrium. The red line represents the static magnetic
field direction, and the black line the Bloch vector $\boldsymbol{k}$. In this
calculation $\omega_{0}=10$, $\gamma_{\mathrm{D}}=5$,
$\boldsymbol{\hat{s}}=\boldsymbol{\hat{z}}$, $\omega_{\mathrm{s}}^{2}=10$ and
$\tau_{\mathrm{s}}=5$.}%
\label{FigOneSpinTE}%
\end{figure}

The effect of a fluctuating magnetic field is demonstrated by the plot shown
in Fig. \ref{FigOneSpinTE}. The time evolution of $\boldsymbol{k}$ is
evaluated by numerically integrating the modified Schr\"{o}dinger equation
(\ref{MSE}) with added fluctuating magnetic field. The parameters used for the
calculation are listed in the figure caption. The Wiener-Khinchine theorem is
employed to relate the given correlation function (\ref{omega ACF}) to the
power spectrum, which, in turn, is used to derive the variance of Fourier
coefficients of $\omega_{x}$, $\omega_{y}$ and $\omega_{z}$. The variance
values are employed for generating random Fourier coefficients, which in turn,
allow generating the functions $\omega_{x}\left(  t\right)  $, $\omega
_{y}\left(  t\right)  $ and $\omega_{z}\left(  t\right)  $ in a way consistent
with Eq. (\ref{omega ACF}).

To account for the effect of the fluctuating field, a longitudinal relaxation term proportional to
$T_{\mathrm{s}1}^{-1}$ is added to Eq. (\ref{d k par / dt om=z}). Consider the case
where $T_{\mathrm{s}1}^{-1}\ll\gamma_{\mathrm{D}}\ll\left\vert
\boldsymbol{\omega
}\right\vert $ and $\boldsymbol{\hat{s}}=\boldsymbol{\hat{z}}$. For this case
Eq. (\ref{d k par / dt om=z}) has a steady state solution given by
$k_{\parallel}=-1+1/\left(  1+2\gamma_{\mathrm{D}}T_{\mathrm{s}1}\right)  $
[the term proportional to $s_{x}k_{x}+s_{y}k_{y}$ in Eq.
(\ref{d k par / dt om=z}) has been disregarded, since it is assumed that
$\gamma_{\mathrm{D}}\ll\left\vert \boldsymbol{\omega}\right\vert $]. The
corresponding effective temperature $T_{\mathrm{eff}}$ is given by%
\begin{equation}
T_{\mathrm{eff}}=\frac{\hbar\omega_{0}}{2k_{\mathrm{B}}\tanh^{-1}\left(
1-\frac{1}{1+2\gamma_{\mathrm{D}}T_{\mathrm{s}1}}\right)  }\;, \label{eff T}%
\end{equation}
where $k_{\mathrm{B}}$ is the Boltzmann's constant.

\section{Two spins}

\label{SecTwoSpins}

Consider a two spin 1/2 system in a pure state $\left\vert \psi\right\rangle $
given by $\left\vert \psi\right\rangle =a\left\vert ++\right\rangle
+b\left\vert +-\right\rangle +c\left\vert -+\right\rangle +d\left\vert
--\right\rangle $.

\subsection{Single spin angular momentum}

The single-spin angular momentum (in units of $\hbar/2$) vector operators are
denoted by $\boldsymbol{S}_{1}=\left(  S_{1x},S_{1y},S_{1z}\right)  $ and
$\boldsymbol{S}_{2}=\left(  S_{2x},S_{2y},S_{2z}\right)  $, and the total spin
angular momentum is $\boldsymbol{S}=\boldsymbol{S}_{1}+\boldsymbol{S}_{2}%
=\left(  S_{x},S_{y},S_{z}\right)  $. A given single-spin linear operator of
spin 1 (2) is represented by the $4\times4$ matrix $\sigma_{0}\otimes K$
($K\otimes\sigma_{0}$) where $\otimes$ denotes the Kronecker tensor product,
$\sigma_{0}$\ is the $2\times2$ identity matrix, and where $K$ is the
$2\times2$ matrix representation of the given single-spin operator. The matrix
representations of $\boldsymbol{S}_{1}\cdot\boldsymbol{\hat{u}}_{1}$ and
$\boldsymbol{S}_{2}\cdot\boldsymbol{\hat{u}}_{2}$, where
$\boldsymbol{\hat{u}}_{1}=\left(  \sin\theta_{1}\cos\varphi_{1},\sin\theta
_{1}\sin\varphi_{1},\cos\theta_{1}\right)  $ and $\boldsymbol{\hat{u}}_{2}%
=\left(  \sin\theta_{2}\cos\varphi_{2},\sin\theta_{2}\sin\varphi_{2}%
,\cos\theta_{2}\right)  $ are unit vectors, are thus given by [see Eq.
(\ref{Pauli})]%
\begin{align}
&  \boldsymbol{S}_{1}\cdot\boldsymbol{\hat{u}}_{1}\nonumber\\
&  \dot{=}\left(
\begin{array}
[c]{cccc}%
\cos\theta_{1} & 0 & \sin\theta_{1}e^{-i\varphi_{1}} & 0\\
0 & \cos\theta_{1} & 0 & \sin\theta_{1}e^{-i\varphi_{1}}\\
\sin\theta_{1}e^{i\varphi_{1}} & 0 & -\cos\theta_{1} & 0\\
0 & \sin\theta_{1}e^{i\varphi_{1}} & 0 & -\cos\theta_{1}%
\end{array}
\right)  \;,\nonumber\\
&  \label{S1*u1}%
\end{align}
and%
\begin{align}
&  \boldsymbol{S}_{2}\cdot\boldsymbol{\hat{u}}_{2}\nonumber\\
&  \dot{=}\left(
\begin{array}
[c]{cccc}%
\cos\theta_{2} & \sin\theta_{2}e^{-i\varphi_{2}} & 0 & 0\\
\sin\theta_{2}e^{i\varphi_{2}} & -\cos\theta_{2} & 0 & 0\\
0 & 0 & \cos\theta_{2} & \sin\theta_{2}e^{-i\varphi_{2}}\\
0 & 0 & \sin\theta_{2}e^{i\varphi_{2}} & -\cos\theta_{2}%
\end{array}
\right)  \;.\nonumber\\
&  \label{S2*u2}%
\end{align}
With the help of Eqs. (\ref{S1*u1}) and (\ref{S2*u2}) and the normalization
condition $aa^{\ast}+bb^{\ast}+cc^{\ast}+dd^{\ast}=1$ one finds that
\begin{equation}
\left\vert \left\langle \boldsymbol{S}_{1}\right\rangle \right\vert
^{2}=\left\vert \left\langle \boldsymbol{S}_{2}\right\rangle \right\vert
^{2}=1-4\left\vert ad-bc\right\vert ^{2}\;. \label{S1^2=S2^2}%
\end{equation}
Note that the normalization condition implies that $\left\vert
ad-bc\right\vert ^{2}\leq1/4$. In standard quantum mechanics the term $ad-bc$
is time-independent, provided that the spins are decoupled [see Eq. (8.121) of
Ref. \cite{Buks_QMLN}]. The term $\left\vert ad-bc\right\vert $ can be
extracted from the partial transpose $\rho^{\mathrm{T}1}$ ($\rho^{\mathrm{T}%
2}$) of the spin-spin density operator with respect to spin 1 (2) using the
relation $\det\rho^{\mathrm{T}1}=\det\rho^{\mathrm{T}2}=-\left\vert
ad-bc\right\vert ^{4}$ \cite{Peres_1413}.

Consider the case where $\left\langle \boldsymbol{S}_{1}\right\rangle
=\left\langle \boldsymbol{S}_{2}\right\rangle =0$. For this case the following
holds $\left\langle S_{1z}\right\rangle =a^{\ast}a+b^{\ast}b-c^{\ast}%
c-d^{\ast}d=0$, $\left\langle S_{2z}\right\rangle =a^{\ast}a-b^{\ast}%
b+c^{\ast}c-d^{\ast}d=0$, $\left\langle S_{1+}\right\rangle =2\left(  a^{\ast
}c+b^{\ast}d\right)  =0$ and $\left\langle S_{2+}\right\rangle =2\left(
a^{\ast}b+c^{\ast}d\right)  =0$, where $S_{n\pm}=S_{nx}\pm iS_{ny}$ and
$n\in\left\{  1,2\right\}  $. The conditions $\left\langle S_{1z}\right\rangle
=\left\langle S_{2z}\right\rangle =0$ imply that $a^{\ast}a=d^{\ast}d$ and
$b^{\ast}b=c^{\ast}c$, whereas the conditions $\left\langle S_{1+}%
\right\rangle =\left\langle S_{2+}\right\rangle =0$ yield $a^{\ast}%
/d=-b^{\ast}/c=-c^{\ast}/b$. Hence the state vector $\left\vert \psi
\right\rangle $ for this case has the form%
\begin{align}
\left\vert \psi\right\rangle  &  =\frac{\cos\frac{\theta_{\psi}}{2}%
e^{-\frac{i\phi_{\alpha}}{2}}}{\sqrt{2}}\left\vert ++\right\rangle +\frac
{\sin\frac{\theta_{\psi}}{2}ie^{-\frac{i\phi_{\beta}}{2}}}{\sqrt{2}}\left\vert
+-\right\rangle \nonumber\\
&  +\frac{\sin\frac{\theta_{\psi}}{2}ie^{\frac{i\phi_{\beta}}{2}}}{\sqrt{2}%
}\left\vert -+\right\rangle +\frac{\cos\frac{\theta_{\psi}}{2}e^{\frac
{i\phi_{\alpha}}{2}}}{\sqrt{2}}\left\vert --\right\rangle \;,\nonumber\\
&  \label{psi <S1>=<S2>=0}%
\end{align}
where $\theta_{\psi}$, $\phi_{\alpha}$ and $\phi_{\beta}$ and are real. Note
that $ad-bc=1/2$ for the state (\ref{psi <S1>=<S2>=0}).

The operator $R$ is defined by [note that $\boldsymbol{S}_{1}\cdot
\boldsymbol{S}_{2}=\boldsymbol{S}_{2}\cdot\boldsymbol{S}_{1}$, see Eqs.
(\ref{S1*u1}) and (\ref{S2*u2})]%
\begin{equation}
R=\boldsymbol{S}_{1}\cdot\boldsymbol{S}_{2}-\left\langle \boldsymbol{S}_{1}%
\right\rangle \cdot\left\langle \boldsymbol{S}_{2}\right\rangle \;, \label{R=}%
\end{equation}
and its expectation value $\left\langle R\right\rangle $ is given by%
\begin{equation}
\left\langle R\right\rangle =4\left(  \left\vert ad\right\vert ^{2}-\left\vert
bc\right\vert ^{2}\right)  -4\operatorname{Re}\left(  \left(  b^{\ast}{}%
^{2}+c^{\ast}{}^{2}\right)  \left(  ad-bc\right)  \right)  \;.
\end{equation}
Note that $\left\langle R\right\rangle =0$ when $ad-bc=0$.

\subsection{Single spin purity}

The single-spin purity $P$ is given by $P=1-2\left\vert ad-bc\right\vert ^{2}$
[see Eq. (8.642) of Ref. \cite{Buks_QMLN}]. It is bounded by $1/2\leq P\leq1$
[recall that $\left\vert ad-bc\right\vert ^{2}\leq1/4$]. In terms of the
purity $P$, Eq. (\ref{S1^2=S2^2}) reads $\left\vert \left\langle
\boldsymbol{S}_{1}\right\rangle \right\vert ^{2}=\left\vert \left\langle
\boldsymbol{S}_{2}\right\rangle \right\vert ^{2}=2\left(  P-1/2\right)  $.

\subsection{Spin-spin disentanglement}

Spin-spin disentanglement is generated by the term proportional to
$\gamma_{\mathrm{D}}$ in the modified Schr\"{o}dinger equation (\ref{MSE})
provided that the bra vector $\left\langle \Psi\right\vert $ is taken to be
given by \cite{Buks_355303,Wootters_2245}%
\begin{equation}
\left\langle \Psi\right\vert =d\left\langle ++\right\vert -c\left\langle
+-\right\vert -b\left\langle -+\right\vert +a\left\langle --\right\vert \;.
\label{bra Psi}%
\end{equation}
Note that $\left\langle \Psi\right\vert $ (\ref{bra Psi}) is normalized
provided that $\left\vert \psi\right\rangle $ is normalized, and that
$\left\langle \Psi\right.  \left\vert \psi\right\rangle =2\left(
ad-bc\right)  $ [compare with Eq. (\ref{S1^2=S2^2})].

\begin{figure}[b]
\begin{center}
\includegraphics[width=1.0\columnwidth]{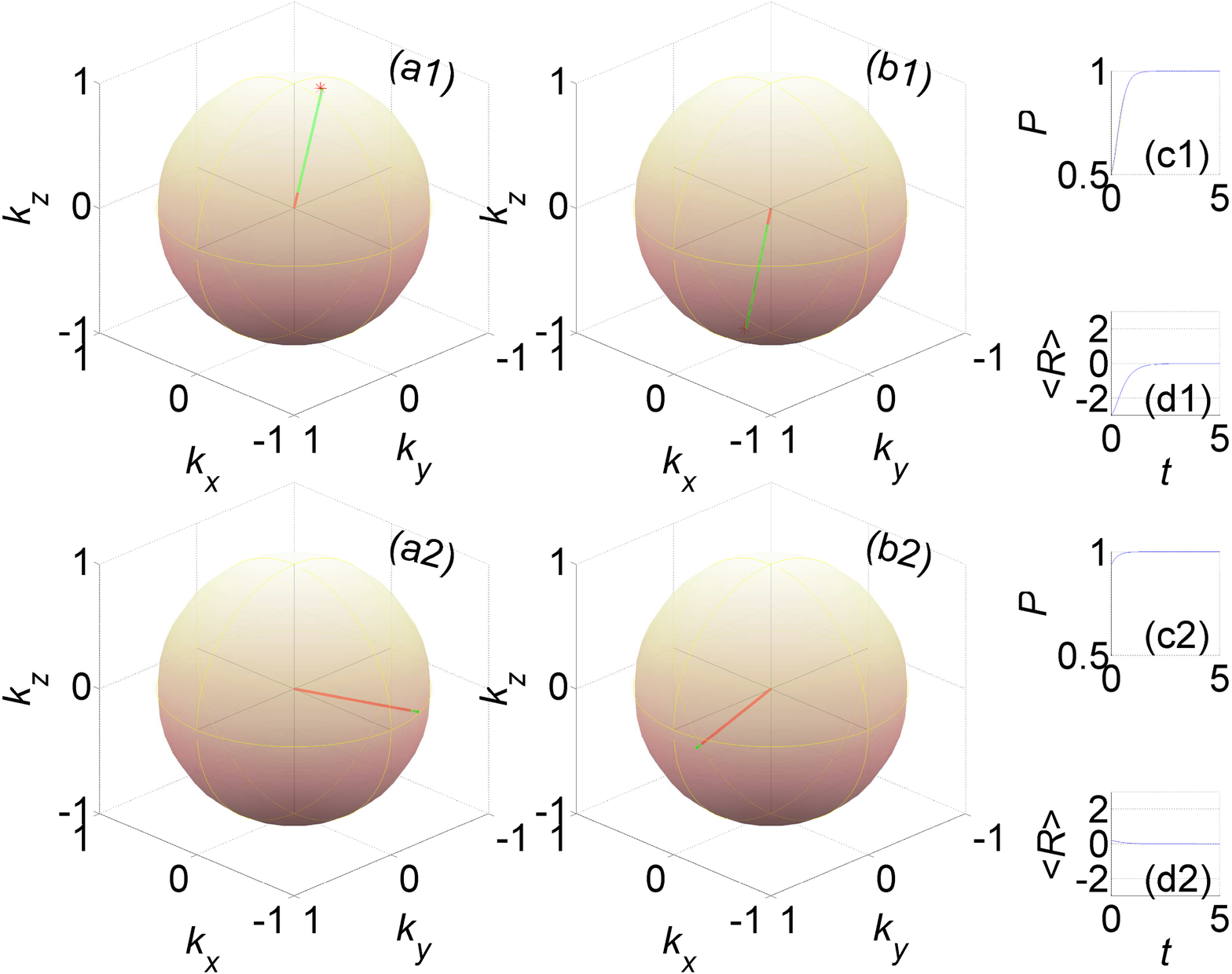}
\end{center}
\caption{Spin-spin disentanglement. In the panels labelled by the numbers '1'
('2'), the single-spin purity is initially $P\simeq1/2$ ($P\simeq1$). The
(initial value) time evolution of $\left\langle \boldsymbol{S}_{1}%
\right\rangle $ and $\left\langle \boldsymbol{S}_{2}\right\rangle $ is
indicated by red (green) lines in the plots labelled by the letter 'a' and
'b', respectively. The single spin purity $P$ and $\left\langle R\right\rangle
$ are shown as a function of time $t$ in the plots labelled by the letters 'c'
and 'd', respectively.}%
\label{FigTwoSpinD}%
\end{figure}

The plots shown in Fig. \ref{FigTwoSpinD} are based on numerical integration
of the spin-spin modified Schr\"{o}dinger equation (\ref{MSE}) with
$\mathcal{H}=0$, and with $\left\langle \Psi\right\vert $ given by Eq.
(\ref{bra Psi}). In the panels labelled by the number '1' ('2'), the
single-spin purity is initially low $P\simeq1/2$ (high $P\simeq1$), i.e.
initially$\ \left\vert ad-bc\right\vert ^{2}\simeq1/4$\ ($\left\vert
ad-bc\right\vert ^{2}\simeq0$). The three-dimensional plots labelled by the
letter 'a' ('b') display the time evolution of $\left\langle
\boldsymbol{S}_{1}\right\rangle $ ($\left\langle \boldsymbol{S}_{2}%
\right\rangle $). In these plots, red straight lines are drawn between the
origin and the initial value of $\left\langle \boldsymbol{S}_{n}\right\rangle
$, whereas green lines represent time evolution of $\left\langle
\boldsymbol{S}_{n}\right\rangle $, where $n\in\left\{  1,2\right\}  $. The
plots labelled by the letters 'c' and 'd' display the time evolution of $P$
and $\left\langle R\right\rangle $ [see Eq. (\ref{R=})], respectively. For
both cases '1' and '2', $\left\vert \left\langle \boldsymbol{S}_{1}%
\right\rangle \right\vert ^{2}=\left\vert \left\langle \boldsymbol{S}_{2}%
\right\rangle \right\vert ^{2}\rightarrow1$ [i.e. $\left\vert ad-bc\right\vert
^{2}\rightarrow0$ and $P\rightarrow1$, see Eq. (\ref{S1^2=S2^2})] in the long
time limit $t\rightarrow\infty$, i.e. entanglement vanishes in this limit.

The case where initially $P\simeq1/2$ [see Fig. \ref{FigTwoSpinD} (a1), (b1),
(c1) and (d1)] is explored by numerically integrating the modified
Schr\"{o}dinger equation (\ref{MSE}) with the initial condition (before
normalization) of $\left\vert \psi\right\rangle =\left\vert \psi
_{0}\right\rangle +\epsilon\left\vert \psi_{\mathrm{p}}\right\rangle $, where
$\left\vert \psi_{0}\right\rangle =\left\vert \mathrm{B}_{0,0}\right\rangle
=2^{-1/2}\left(  \left\vert +-\right\rangle -\left\vert -+\right\rangle
\right)  $ is the singlet Bell state (which is invariant under single-spin
basis transformation, and which is fully entangled), $\epsilon$ is a small
positive number, and $\left\vert \psi_{\mathrm{p}}\right\rangle =\alpha
\left\vert ++\right\rangle +\beta\left\vert +-\right\rangle +\gamma\left\vert
-+\right\rangle +\delta\left\vert --\right\rangle $ is normalized. For this
case, to first order in $\epsilon$ initially $\left\langle S_{1+}\right\rangle
=-\left\langle S_{2+}\right\rangle =2^{1/2}\left(  \delta-\alpha^{\ast
}\right)  \epsilon$ and $\left\langle S_{1z}\right\rangle =-\left\langle
S_{2z}\right\rangle =2^{-1/2}\left(  \beta+\beta^{\ast}+\gamma+\gamma^{\ast
}\right)  \epsilon$, i.e. initially $\left\langle \boldsymbol{S}_{1}%
\right\rangle =-\left\langle \boldsymbol{S}_{2}\right\rangle $. The plots in
Fig. \ref{FigTwoSpinD} (a1), (b1), (c1) and (d1) demonstrate that both
$\left\langle \boldsymbol{S}_{1}\right\rangle $ and $\left\langle
\boldsymbol{S}_{2}\right\rangle $ increase in magnitude with time while
remaining nearly anti-parallel to each other as they both approach the Bloch
sphere surfaces. The red star symbols in Fig. \ref{FigTwoSpinD}(a1) and (b1)
indicate the initial values of $\left\langle \boldsymbol{S}_{1}\right\rangle
/\left\vert \left\langle \boldsymbol{S}_{1}\right\rangle \right\vert $ and
$\left\langle \boldsymbol{S}_{2}\right\rangle /\left\vert \left\langle
\boldsymbol{S}_{2}\right\rangle \right\vert $. As the plots in Fig.
\ref{FigTwoSpinD}(a1) and (b1) demonstrate, time evolution leaves these
normalized values nearly unchanged, provided that $\epsilon\ll1$ (the value of
$\epsilon=0.1$ has been used for generating the plots).

The case shown in Fig. \ref{FigTwoSpinD} (a1), (b1), (c1) and (d1)
demonstrates strong dependency of the long time value of $\left\vert
\psi\right\rangle $ on its initial value. This dependency, which becomes
extreme when $\left\vert \psi\right\rangle $ is initially fully entangled,
resembles the butterfly effect. In the limit $\epsilon\rightarrow0$, angular
momentum is conserved by the modified Schr\"{o}dinger equation, provided that
$\left\vert \psi_{0}\right\rangle =\left\vert \mathrm{B}_{0,0}\right\rangle $.
This can be attributed to the fact that $S_{x}\left\vert \mathrm{B}%
_{0,0}\right\rangle =S_{y}\left\vert \mathrm{B}_{0,0}\right\rangle
=S_{z}\left\vert \mathrm{B}_{0,0}\right\rangle =0$. Conservation of the
angular momentum $\boldsymbol{\hat{z}}$ component is obtained when the initial
state is the Bell triplet state $\left\vert \mathrm{B}_{1,0}\right\rangle
=2^{-1/2}\left(  \left\vert +-\right\rangle +\left\vert -+\right\rangle
\right)  $, which is also fully entangled, and for which $S_{z}\left\vert
\mathrm{B}_{1,0}\right\rangle =0$.

The case where initially $P\simeq1$ is demonstrated by Fig. \ref{FigTwoSpinD}
(a2), (b2), (c2) and (d2). For this case both $\left\langle \boldsymbol{S}_{1}%
\right\rangle $ and $\left\langle \boldsymbol{S}_{2}\right\rangle $ are
initially close to the Bloch sphere surfaces. Consequently, the time evolution
of $\left\vert \psi\right\rangle $ towards a fully product state does not
significantly change $\left\langle \boldsymbol{S}_{1}\right\rangle $ and
$\left\langle \boldsymbol{S}_{2}\right\rangle $ (time evolution is represented
by the green lines).

\section{Instability}

\label{SecInstab}

Consider a system composed of two spins 1/2. The first spin, which is labelled
as '$\mathrm{a}$', has a relatively low angular frequency $\omega_{\mathrm{a}%
}$ in comparison with the angular frequency $\omega_{\mathrm{b}}$ of the
second spin, which is labelled as '$\mathrm{b}$', and which is externally
driven. The angular momentum vector operator of particle $\mathrm{a}$
($\mathrm{b}$) is denoted by $\boldsymbol{S}_{\mathrm{a}}$
($\boldsymbol{S}_{\mathrm{b}}$). The Hamiltonian $\mathcal{H}$ of the closed
system is given by%
\begin{equation}
\mathcal{H}=\omega_{\mathrm{a}}S_{\mathrm{az}}+\omega_{\mathrm{b}%
}S_{\mathrm{bz}}+\frac{\omega_{1}\left(  S_{\mathrm{b+}}+S_{\mathrm{b-}%
}\right)  }{2}+V\;, \label{H TS}%
\end{equation}
where the driving amplitude and angular frequency are denoted by $\omega_{1}$
and $-\omega_{\mathrm{p}}=\omega_{\mathrm{b}}-\Delta$, respectively ($-\Delta$
is the driving detuning), the operators $S_{\mathrm{a\pm}}$ are given by
$S_{\mathrm{a\pm}}=S_{\mathrm{ax}}\pm iS_{\mathrm{ay}}$, and the rotated
operators $S_{\mathrm{b\pm}}$ are given by $S_{\mathrm{b\pm}}=\left(
S_{\mathrm{bx}}\pm iS_{\mathrm{by}}\right)  e^{\mp i\omega_{\mathrm{p}}t}$.
The coupling term is given by $V=g\hbar^{-1}\left(  S_{\mathrm{a+}%
}+S_{\mathrm{a-}}\right)  S_{\mathrm{bz}}$, where $g$ is a coupling rate. In a
rotating frame, the matrix representation of the transformed Hamiltonian
$\mathcal{H}^{\prime}$ is given by $\mathcal{H}^{\prime}\dot{=}\hbar\Omega$,
where the $4\times4$ matrix $\Omega$ is given by%
\begin{equation}
\Omega=\left(
\begin{array}
[c]{cccc}%
\frac{\omega_{\mathrm{a}}+\Delta}{2} & \frac{\omega_{1}}{2} & \frac{g}{2} &
0\\
\frac{\omega_{1}}{2} & \frac{\omega_{\mathrm{a}}-\Delta}{2} & 0 & -\frac{g}%
{2}\\
\frac{g}{2} & 0 & \frac{-\omega_{\mathrm{a}}+\Delta}{2} & \frac{\omega_{1}}%
{2}\\
0 & -\frac{g}{2} & \frac{\omega_{1}}{2} & \frac{-\omega_{\mathrm{a}}-\Delta
}{2}%
\end{array}
\right)  \;.
\end{equation}

\begin{figure}[b]
\begin{center}
\includegraphics[width=1.0\columnwidth]{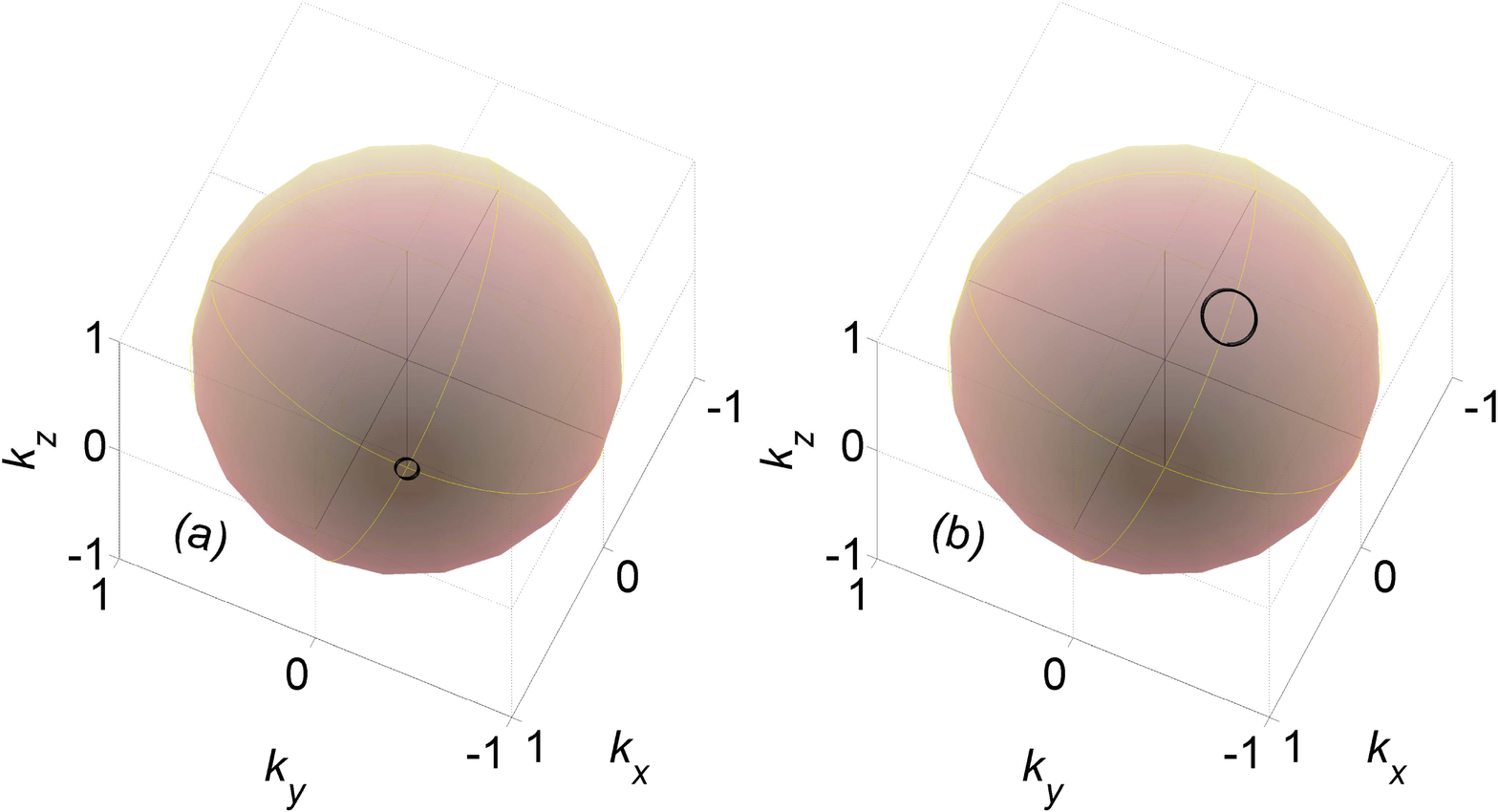}
\end{center}
\caption{Dipolar LC. The expectation values of $\left\langle
\boldsymbol{S}_{1}\right\rangle $ and $\left\langle \boldsymbol{S}_{2}%
\right\rangle $ are shown in (a) and (b), respectively. The modified
Schr\"{o}dinger equation (\ref{MSE}), with the Hamiltonian (\ref{H TS}) and
the bra vector (\ref{bra Psi}) is numerically integrated. The parameters used
for the calculation are $\gamma_{\mathrm{D}}=10^{3}$, $\omega_{\mathrm{a}%
}=10^{2}$, $\omega_{1}=-\Delta=2^{-1/2}\omega_{\mathrm{a}}$ and $g=0.2$.
Fluctuating magnetic field with parameters $\omega_{\mathrm{s}}^{2}=1$ and
$\tau_{\mathrm{s}}=0.05$ is added [see Eq. (\ref{omega ACF})].}%
\label{FigTwoSpinLC}%
\end{figure}

Disentanglement is generated by the modified Schr\"{o}dinger equation
(\ref{MSE}) by choosing the bra vector $\left\langle \Psi\right\vert $ to be
given by Eq. (\ref{bra Psi}). The expectation values of $\left\langle
\boldsymbol{S}_{1}\right\rangle $ and $\left\langle \boldsymbol{S}_{2}%
\right\rangle $, which are shown in Fig. \ref{FigTwoSpinLC}(a) and in Fig.
\ref{FigTwoSpinLC}(b), respectively, are calculated by numerically integrating
the modified Schr\"{o}dinger equation (\ref{MSE}). For the plot shown in Fig.
\ref{FigTwoSpinLC}, for which the Hartmann--Hahn matching condition
$\omega_{\mathrm{a}}=\omega_{\mathrm{R}}$
\cite{Hartmann1962,Yang_1,Levi_053516}\ is assumed to be satisfied, where
$\omega_{\mathrm{R}}=\sqrt{\omega_{1}^{2}+\Delta^{2}}$ is the Rabi angular
frequency, both $\left\langle \boldsymbol{S}_{1}\right\rangle $ and
$\left\langle \boldsymbol{S}_{2}\right\rangle $ undergo a limit cycle (LC).
The instability responsible for the LC was studied in Ref. \cite{Levi_053516},
in which the equations of motion generated by the Hamiltonian (\ref{H TS})
were treated using in the mean-field approximation. This example demonstrates
the connection between the mean-field approximation and disentanglement.

\section{Summary}

In summary, both processes of thermalization and disentanglement can be
modeled using a recently proposed modified Schr\"{o}dinger equation. The added
nonlinear term can give rise to instabilities and LC solutions. On
the other hand, it remains unclear whether quantum mechanics can be
self-consistently reformulated based on the proposed modified Schr\"{o}dinger
equation. Future study will be devoted to exploring candidate formalisms.

\section{Acknowledgments}

This work was supported by the Israeli science foundation, the Israeli ministry of science, and by the Technion security research foundation. We thank Michael R. Geller for helpful discussions.

\bibliographystyle{ieeepes}
\bibliography{acompat,Eyal_Bib}

\end{document}